\documentclass[a4paper,fleqn]{cas-dc}

\usepackage[numbers]{natbib}

\usepackage[nohyperlinks,nolist]{acronym}
\begin{acronym}
	\acro{BCI}{brain--computer interface}
	\acro{SNR}{signal-to-noise ratio}
	\acro{EEG}{electroencephalography}
	\acro{SATO}{speed--accuracy trade-off}
	\acro{ERP}{event-related potential}
	\acro{SSVEP}{steady-state visual evoked potential}
	\acro{SCP}{slow-cortical potential}

	\acro{RSVP}{rapid serial visual presentation}
	\acro{RCP}{row--column paradigm}
	\acro{RRCP}{stimulus-level RCP}

	\acro{ISI}{inter-stimulus interval}
	\acro{SOA}{stimulus onset asynchrony}
	
	\acro{GCB}{Gain--Cons Balance}

	\acro{CV}{cross-validation}
	\acro{LOSSO}{leave-one-session-out}

	\acro{SVC}{Support Vector Classifier}
	\acro{RLDA}{Regularized Linear Discriminant Analysis}
	\acro{RF}{Random Forest}
	\acro{PR-AUC}{precision-recall area under the curve}

	\acro{ESS}{early-stopping strategy}
	\acroplural{ESS}{early-stopping strategies}
	\acro{FS}{fixed-stop}
	\acro{AE}{accumulated evidence}
	\acro{ST}{statistical test}

	\acro{BCIm}{BCI measurement}
	\acro{ITR}{Information Transfer Rate}
	\acro{bBCI}{baseline BCI}
	\acro{ROI}{region of interest}
\end{acronym}

\begin{document}
\let\WriteBookmarks\relax
\def\floatpagepagefraction{1}
\def\textpagefraction{.001}

\shorttitle{Framework to Control Speed--Accuracy Trade-Off in BCIs}

\title [mode = title]{A Methodological Framework for Explicit Control of the Speed--Accuracy Trade-off in Brain--Computer Interfaces}

\shortauthors{J. Jiménez et~al.}

\affiliation[1]{organization={Grupo de Neurocomputación Biológica, Departamento de Ingeniería Informática, Universidad Autónoma de Madrid},
                city={Madrid},
                postcode={28049}, 
                country={Spain}}

\author[1]{J. Jiménez}[orcid=0009-0009-2483-7124]
\cormark[1]
\ead{javier.jimenez01@uam.es}

\credit{Conceptualization, Methodology, Software, Writing - original draft}

\author[1]{F.B. Rodríguez}[orcid=0000-0003-4053-099X]
\cormark[1]
\ead{f.rodriguez@uam.es}

\credit{Conceptualization, Methodology, Writing - review \& editing}

\cortext[cor1]{Corresponding author}


\begin{abstract}
Current \acp{BCI} are limited by low \acl{SNR} in modalities such as \acl{EEG}, which requires multiple trials to reliably decode user intentions.
This induces an inherent \acl{SATO}, whereby higher accuracy comes at the cost of speed.
The acceptable speed--accuracy balance is application-dependent, motivating explicit and user-controllable trade-offs.
Conventional metrics, such as the \acl{ITR}, combine speed and accuracy obscuring their dependence and potentially introducing systematic biases.

In this study, we propose an evaluation framework independent of classifier, paradigm, and \acl{ESS} that explicitly separates and optimizes speed and accuracy.
We employ two measures from prior work, \emph{Gain} (relative speed improvement) and \emph{Conservation} (relative accuracy preservation), and combine them into a tunable \acl{GCB} controlled by $\alpha$, explicitly regulating the \acl{SATO}.
The parameter continuously adjusts the operating point without modifying the underlying classifier, facilitating deployment across scenarios.

The framework was evaluated on P300 \acl{ERP} paradigms using public recordings from 63 subjects as well as multiple classifiers and \aclp{ESS} to achieve distinct operating points in speed--accuracy and bitrate.
Results show that tuning $\alpha$ yields fast, accurate, or balanced \ac{BCI} behaviours, demonstrating explicit control of the \acl{SATO}.
The method supports subject-level performance prediction and improves explainability of \ac{BCI} behaviour.
Further analysis of the \acl{ITR} reveals a systematic bias toward speed, explained by the proposed framework through the \emph{Gain} and \emph{Conservation} measurements.
Overall, this work establishes the \acl{SATO} as a controllable design variable validated on public P300-based paradigms, enabling transparent evaluation and application-specific optimization of \acp{BCI}.
\end{abstract}

\begin{graphicalabstract}
  \begin{center}
    \includegraphics[width=\textwidth]{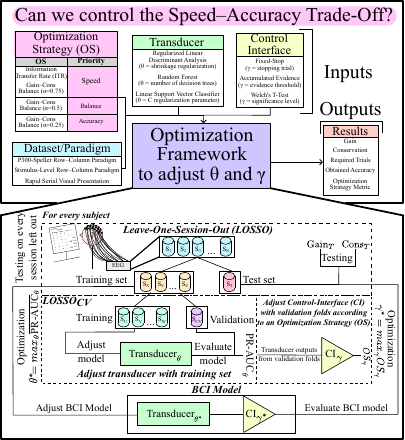}
  \end{center}
\end{graphicalabstract}

\begin{highlights}
  \item Proposes the Gain--Cons framework to decouple and control speed and accuracy in BCIs.
  \item Transforms the \acl{SATO} into a controllable variable via parameter $\alpha$.
  \item Introduces a procedure to jointly optimize transducers and control-interfaces.
  \item Maps speed--accuracy trade-offs to predict required trials and expected accuracy.
  \item Shows that the \acl{ITR} exhibits an inherent bias toward speed.
\end{highlights}



\maketitle

\acresetall
\section{Introduction}
\begin{figure}[pos=h]
	\includegraphics[width=\linewidth]{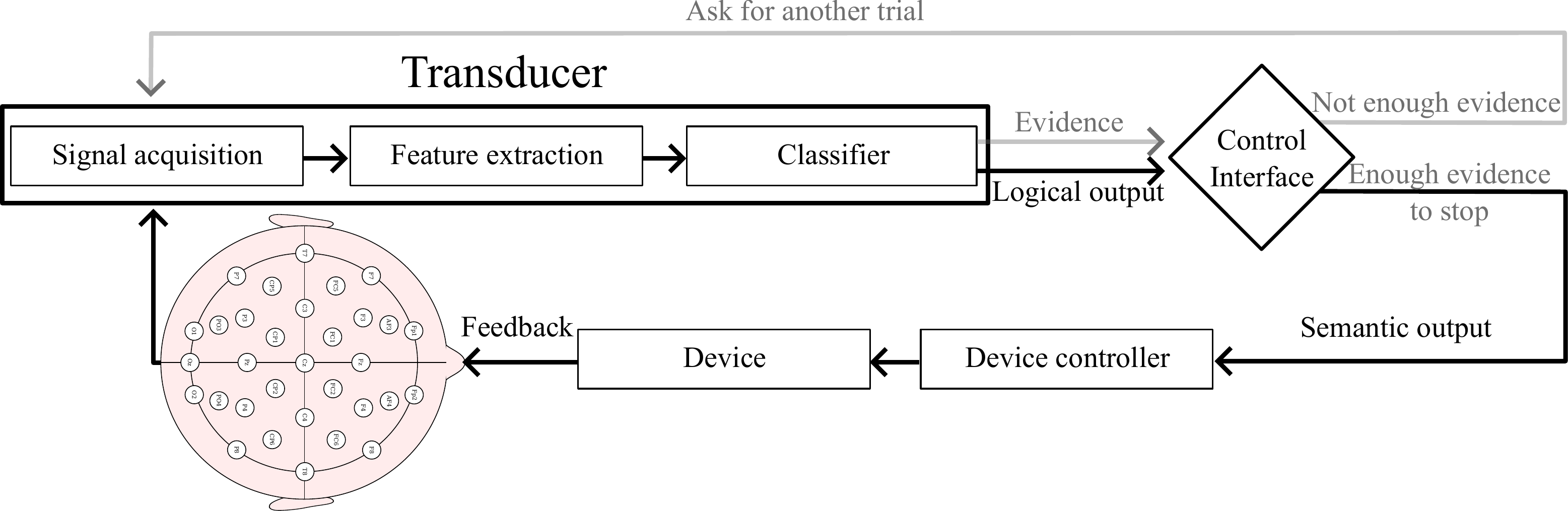}
	\caption{
		Closed-loop \ac{BCI} system composed of a transducer, a control-interface, and a device controller.
		A grey line highlights \acp{BCI} can stop sending commands if not enough evidence is available.
		Inspired from \cite{bianchi_performances_2007,mason_general_2003}.
		This diagram provides a conceptual framework for the separation of signal decoding and decision control.
	}
	\label{fig:BCI_Sys}
\end{figure}

\Acp{BCI} decode brain signals into commands for external devices \cite{hoffmann_efficient_2008,won_eeg_2022}.
They are typically composed of a transducer, which translates \ac{EEG} signals into logical outputs, a control-interface that assigns semantic meaning to these outputs, and a device controller that executes commands and provides user feedback \cite{bianchi_performances_2007,mason_general_2003} as illustrated in Fig.~\ref{fig:BCI_Sys}.

To decode raw \ac{EEG} into commands to be executed by external devices, \ac{BCI} designers commonly employ \acp{ERP} as control signals, which are time-locked physiological responses that can be extracted through spatial filtering and pattern recognition methods \cite{edelman_non-invasive_2025}.
Among the most widely used exogenous control signals are \acp{SSVEP}, \aclp{SCP}, and P300-\acp{ERP} \cite{ramadan_brain_2017}.
Our work focuses on the P300-\ac{ERP}, a positive deflection in the \ac{EEG} signal occurring approximately 300~ms after the presentation of an infrequent and task-relevant stimulus \cite{polich_updating_2007}.

The P300-\ac{ERP} is typically elicited using the oddball paradigm, in which rare target stimuli are interspersed among frequent non-target stimuli to direct the user’s attention.
Common implementations of this paradigm include the \ac{RSVP}, where stimuli are presented sequentially at high speed in a single spatial location \cite{lees_review_2018,hoffmann_efficient_2008}, and the \ac{RCP}, where rows and columns of a character grid are intensified while the user focuses on a target symbol \cite{won_eeg_2022}.
Both paradigms were considered in this study to analyze their impact on P300-\ac{ERP} detection performance.

Despite recent advances in signal processing and machine learning \cite{lotte_review_2018,edelman_non-invasive_2025}, \ac{BCI} performance remains constrained by the low \ac{SNR} of commonly used modalities such as \ac{EEG}.
This limitation is particularly critical in \ac{ERP}-based paradigms, such as P300 detection tasks, where task-relevant signals can be difficult to distinguish from background activity.

A common strategy to improve reliability is trial averaging over repeated time-locked events, which enhances signal detectability by increasing the effective \ac{SNR} \cite{luck_chapter_2014}.
However, this approach introduces a fundamental \acl{SATO}, whereby increasing the number of trials improves accuracy at the cost of reduced communication speed \cite{schreuder_optimizing_2013}.
Crucially, the acceptable balance between speed and accuracy depends on the application, as some tolerate occasional errors (e.g., \ac{BCI} spellers) while others require high reliability (e.g., wheelchair control) \cite{bianchi_performances_2007}.
This means that the same \ac{BCI} system may require different configurations depending on the context of use or the user.

Addressing this trade-off requires selecting an appropriate stopping point, which is commonly formalized through \aclp{BCIm} that combine speed and accuracy into a single value, such as the \ac{ITR} \cite{wolpaw_braincomputer_2002}, BCI-Utility \cite{dal_seno_utility_2010}, or Efficiency \cite{bianchi_performances_2007}.
However, these metrics obscure how accuracy depends on the number of trials and may introduce implicit biases toward speed or accuracy, thereby limiting explicit, interpretable, and application-driven control of the trade-off.
These metrics implicitly enforce a balance between speed and accuracy that neither the user nor the application can directly modify.

Motivated by this limitation, we propose a framework that monitors and optimizes speed and accuracy in an interpretable manner with the speed \emph{Gain} and accuracy \emph{Conservation} measurements introduced in previous work \cite{jimenez_novel_nodate}.
Unlike previous approaches, the proposed framework does not combine speed and accuracy into a single metric, but rather treats them as independent and explicitly controllable dimensions.
This allows the system's operating point to be explicitly defined, rather than having it determined implicitly by the metric used.
The framework does not define a new classification algorithm, but rather a systematic criterion for regulating the operating point of a \ac{BCI} in terms of speed and accuracy.
Our central hypothesis is that decoupling these two dimensions enables more transparent evaluation, improves application-driven control of the \acl{SATO}, and reveals biases inherent in conventional \aclp{BCIm}.
Specifically, we investigate whether this strategy improves trade-off control, supports subject-level performance prediction, and explains the behaviour of standard \aclp{BCIm}.
We validate our hypotheses by means of two \ac{EEG} datasets with different paradigms (\ac{RSVP} and \ac{RCP}), three distinct classifiers as transducers, and three \aclp{ESS} as control-interfaces.
The systematic combination of these elements makes it possible to analyze how different system configurations influence the \acl{SATO}.

The objectives of this study are
(i) to design an optimization framework for evaluating \acp{BCI} with respect to the \acl{SATO},
(ii) to validate it using \ac{EEG} datasets across different paradigms,
(iii) to evaluate combinations of transducers and control-interfaces,
(iv) to analyze population-level performance for robustness and predictive capacity, and
(v) to compare the proposed framework against standard \aclp{BCIm} in terms of performance and bias.
Overall, this work introduces a general and explainable framework to control the \acl{SATO} in \acp{BCI}, enabling fine-grained and application-specific optimization across paradigms, users, and system configurations.

Finally, Section \ref{sec:sota} discusses related work, Section \ref{sec:meth} details the methods and datasets, Section \ref{sec:ress} presents the results, Section \ref{sec:disc} discusses the implications of this work, and Section \ref{sec:conc} concludes the study.

\section{Related Work} \label{sec:sota}
This section reviews existing approaches to evaluating and optimizing performance in \acp{BCI}, with the aim of contextualizing the limitations identified in the introduction and justifying the need for explicit control of the \acl{SATO}.

\subsection{Trends in Measuring BCI Performance}
Most \ac{BCI} \ac{ERP} detection research has focused either on improving signal processing and classification performance or on maximizing global performance metrics \cite{lotte_review_2018,yuan_study_2013,ramsey_brain-computer_2020,yadav_electroencephalogram_2023}.
In \ac{ERP}-based paradigms, studies commonly report measures such as accuracy or Cohen's kappa \cite{cohen_coefficient_1960,lotte_review_2018}.
While these metrics quantify the discriminative performance of transducers, they do not account for communication speed and therefore provide a partial view of \ac{BCI} usability.

To address the \acl{SATO}, composite \aclp{BCIm} have been proposed to integrate speed and accuracy into a single value.
The most widely used metric is the \ac{ITR} \cite{wolpaw_braincomputer_2002}, which combines classification accuracy, number of commands, and decision time to estimate bitrate.

Several variants of the \ac{ITR} described by Wolpaw et al.~\cite{wolpaw_braincomputer_2002} have been proposed to relax its assumptions or adapt it to specific paradigms \cite{arslan_information_2024,bulhoes_da_silva_costa_channel_2020,shi_estimating_2024}.
Other information-theoretic measures, such as Mutual Information Rate, extend this formulation by accommodating unequal prior probabilities of stimuli \cite{wang_asynchronous_2024}.

Beyond information-theoretic approaches, alternative metrics such as BCI-Utility \cite{dal_seno_utility_2010,ma_bci-utility_2023} and Efficiency \cite{bianchi_performances_2007,quitadamo_evaluation_2012} incorporate system constraints and error costs.
At the application level, measures such as Average Time Consumption per Character \cite{wang_average_2024} and Symbols Per Minute \cite{schreuder_optimizing_2013} provide a more direct assessment of practical communication performance.
These metrics are often used as broad indicators of performance, but they do not explicitly explain how that performance is achieved.
This analysis highlights the lack of explicit mechanisms for balancing speed and accuracy, which directly motivates the framework proposed in this paper.

\subsection{Limitations of BCI Performance Metrics}
Despite their widespread use, each \acl{BCIm} relies on specific assumptions.
For example, \ac{ITR} assumes that (i) \acp{BCI} operate as memory-less and stationary discrete communication channels, (ii) output commands are equally likely and classification accuracy is uniform across symbols, and (iii) errors are evenly distributed among non-target classes.
These assumptions are often violated in realistic settings, particularly in early-stopping paradigms or asynchronous \acp{BCI}, where prior probabilities and error distributions are inherently unbalanced \cite{yuan_study_2013}.
Consequently, \ac{ITR} may not accurately reflect practical performance in adaptive or memory-dependent systems.

These limitations become especially evident in early-stopping methods \cite{schreuder_optimizing_2013,ahmadi_bayesian_2024,soetedjo_maintaining_2020}, trial-shortening strategies \cite{gao_improve_2021}, and error-aware correction mechanisms \cite{kalaganis_error-aware_2024}, which explicitly alter temporal or probabilistic structures.
More generally, composite \aclp{BCIm} obscure the relative contributions of speed and accuracy to the final score.

Moreover, Dal Seno et al.~\cite{dal_seno_utility_2010} shown that \ac{ITR} can report non-zero bitrates even when meaningful communication is not achievable, motivating the development of BCI-Utility to ensure performance exceeds chance level given system constraints.
Similarly, Bianchi et al.~\cite{bianchi_performances_2007} highlighted that acceptable error rates are application-dependent, noting that errors tolerated in spellers may be unacceptable in safety-critical systems such as wheelchair control.
Their Efficiency metric incorporates error costs through an Extended Overtime Matrix, emphasizing the importance of context-aware evaluation.

\subsection{Speed--Accuracy Trade-Off Optimization}
Although existing metrics incorporate system constraints \cite{dal_seno_utility_2010} and application-dependent error costs \cite{bianchi_performances_2007}, they still obscure interactions between speed and accuracy.
As a result, they do not explicitly represent preferences between these dimensions nor allow direct control over the \acl{SATO}.
To our knowledge, no prior framework decouples speed and accuracy as independent, tunable dimensions for \ac{BCI} optimization.
In most previous approaches, the trade-off between speed and accuracy is implicitly determined by the chosen metric.
This means that the system's behaviour is determined by the metric used, rather than being directly adjustable by the user or the application.

Several studies have nevertheless addressed the \acl{SATO} through global metrics, early-stopping strategies, and joint optimization of system components.
Early-stopping methods reduce decision time while preserving accuracy by adapting evidence accumulation, as shown in P300 spellers with adaptive \ac{SOA} \cite{gao_improve_2021} and language model integration \cite{soetedjo_maintaining_2020}, as well as in \ac{SSVEP} systems where they mitigate fatigue and reduce selection time \cite{han_enhancing_2024}.
Systematic evaluations further confirm consistent gains in communication speed across methods \cite{schreuder_optimizing_2013}.
Complementarily, cost-sensitive approaches explicitly model error consequences: Bayesian formulations enable control over error types \cite{ahmadi_bayesian_2024}, while the Efficiency metric incorporates both error recovery and time costs through extended confusion and overtime matrices \cite{bianchi_performances_2007,quitadamo_evaluation_2012}, and error-related potentials have been explored to integrate user feedback \cite{kalaganis_error-aware_2024}.

Finally, optimization frameworks have been proposed to jointly tune transducers and control interfaces, either by maximizing speed-prioritizing metrics or by incorporating error costs \cite{schreuder_optimizing_2013,bianchi_performances_2007}.
While effective, these approaches embed the \acl{SATO} within a single objective, limiting interpretability and direct control.
In contrast, our work treats speed and accuracy as independent and controllable variables, enabling explicit adjustment of the \acl{SATO} while retaining a joint optimization perspective.
This approach makes it possible to analyze the effect of each component of the system on speed and accuracy separately.

\section{Methodology} \label{sec:meth}
\subsection{Datasets}
To evaluate the framework under realistic conditions, we defined four dataset inclusion criteria:
(i) recordings acquired with \ac{EEG}, a low-\ac{SNR} modality;
(ii) experiments involving P300-\ac{ERP} detection tasks, which require repeated trials and exhibit a \acl{SATO};
(iii) a \ac{SOA} below 500 ms to ensure fast-paced paradigms; and
(iv) exclusively brain-signal-based interaction, ensuring purely \ac{BCI}-driven communication.
Following these criteria, we selected two public P300 datasets representing distinct paradigms as summarized in Table \ref{tab:Dataset_Attributes}.

The first dataset was extracted from Hoffmann et al.~\cite{hoffmann_efficient_2008}, which implements a \ac{RSVP} paradigm where one out of six visual stimuli is the target.
The second dataset was recorded by Won et al.~\cite{won_eeg_2022}, which follows a \ac{RCP} paradigm where two out of twelve stimuli per trial correspond to the target, representing the row and column of a character.
Both datasets contain recordings acquired with 32 electrodes under the 10--20 system.
Additionally, we derived a third dataset---termed \acl{RRCP}---to remove the conjunctive decoding constraint required for character identification in Won et al.~\cite{won_eeg_2022} \ac{RCP} experiment.
In this case, performance is evaluated at the stimulus level rather than the row--column symbol level, avoiding the logical dependencies of joint decoding and isolating the structural contribution of the paradigm to the \acl{SATO}.

\begin{table}[cols=4,pos=h]
	\centering
	\caption{Properties of the datasets included in the study.}
	\begin{tabular*}{\tblwidth}{LCC}
		\toprule
		\textbf{Property} & \textbf{Hoffmann} & \textbf{Won} \\
		\midrule
		Paradigm & \acs{RSVP} & \acs{RCP} \\
		Stimuli per trial & 6 & 12 \\
		Targets per trial & 1 & 2 (1 row + 1 column) \\
		Trials per run & 20--25 & 15 \\
		Runs per session & 6 & 5--7 \\
		Sessions per subject & 4 & 6 \\
		Subjects & 8 (4 impaired) & 55 \\
		\bottomrule
	\end{tabular*}
	\label{tab:Dataset_Attributes}
\end{table}

\subsection{Preprocessing}
Preprocessing followed the procedures described in the original studies to ensure comparable classification performance.
For Hoffmann et al.~\cite{hoffmann_efficient_2008} \ac{RSVP}, signals were referenced to the average of mastoid electrodes.
A sixth-order Butterworth bandpass filter between 1–12 Hz was applied.
Signals were downsampled from 2048 Hz to 32 Hz via decimation.
One-second epochs were extracted from stimulus onset.
Finally, the number of trials per run was limited to 20.

For both Won et al.~\cite{won_eeg_2022} modalities (\ac{RCP} and \acl{RRCP}), common average referencing was applied.
A fourth-order Butterworth bandpass filter between 0.5–10 Hz was used.
Epochs of 600 ms were extracted from stimulus onset.
Baseline correction was performed using the 200 ms pre-stimulus interval.
Epochs were downsampled from 512 Hz to 20 Hz by averaging non-overlapping samples.
Finally, the number of runs per subject was limited to 5.

\subsection{Transducers and Control-Interfaces}
We selected three probabilistic classifiers capable of producing class membership scores as transducers.
These scores---hereafter referred to as \textit{evidence}---were used to accumulate information across trials within control-interfaces.
In this context, the evidence refers to the probabilities or class membership scores produced by the classifiers for each stimulus.
Specifically, we employed \ac{SVC}, \ac{RLDA}, and \acl{RF} as implemented in Scikit-Learn \cite{pedregosa_scikit-learn_2011}.

Once epoch-level probabilities were obtained, we implemented three \aclp{ESS} as control-interfaces to determine stopping decisions.
Particularly, \acl{FS} stops after a fixed number of trials.
\Acl{AE} aggregates probabilistic outputs across trials until a predefined threshold is reached.
\Acl{ST} terminates trial acquisition once a one-sided t-test with unequal variances detects a significant difference between target and non-target epochs along trials \cite{hohne_two-dimensional_2010}.
All these strategies were chosen due to their demonstrated performance in prior literature \cite{schreuder_optimizing_2013,lotte_review_2018}.

\subsection{BCI Measurements}
To control the \acl{SATO}, we separated speed and accuracy into independent measurements.
Speed was captured by \textit{Gain}, defined as the relative reduction in trials compared to a \acl{bBCI}.
This baseline corresponds to the scenario in which all available trials are used, thereby representing the maximum possible evidence available in a given experiment before a decision is made:
\begin{equation}
	Gain(t, t^*) = \frac{t - t^*}{t},
	\label{eq:gain}
\end{equation}
where $t$ denotes the number of trials required by the \acl{bBCI} which we fix to the maximum number of available trials, and $t^*$ the number of trials required by our control-interface.
It is bounded in [0, 1], representing the proportion of trials saved, with higher values indicating earlier decisions based on less evidence.

Accuracy was encoded through \textit{Conservation} (Cons), defined as the relative preservation of accuracy with respect to the most accurate \acl{bBCI}:
\begin{equation}
	Cons(i, i^*) = 1 - \frac{A(i) - A(i^*)}{A(i)},
	\label{eq:cons}
\end{equation}
where $A(\cdot)$ denotes accuracy of a transducer $\in [0, 1]$ computed as the number of runs correctly classified divided by all available runs after observing a specific number of trials, $i$ the baseline stopping trial which is defined as the latest available trial, and $i^*$ the evaluated stopping point by our control-interface.
Also bounded in [0, 1], higher Conservation indicates greater accuracy retention.

Although dependent on baseline selection, these formulations enable progressive redefinition of the baseline as systems improve.
For comparison, we also computed the \ac{ITR} \cite{wolpaw_braincomputer_2002}.
\begin{gather}
	\begin{aligned}
		B(N,P) &= \log_2 N
		+ P\log_2 P \\
		&\quad + (1-P)\log_2\frac{1-P}{N-1}, \\
		ITR(N,P,\tau) &= B(N,P)\frac{60}{\tau} \quad (\frac{bits}{minute}),
	\end{aligned}
	\label{eq:ITR}
\end{gather}
$B$ represents the number of bits transferred, $\tau$ the time invested in seconds, $N$ the number of possible decisions of the \ac{BCI}, and $P$ the \ac{BCI}'s accuracy.

\subsection{Optimization and Validation Framework} \label{meth:val}
\begin{figure}[pos=h]
	\centering
	\includegraphics[width=\linewidth]{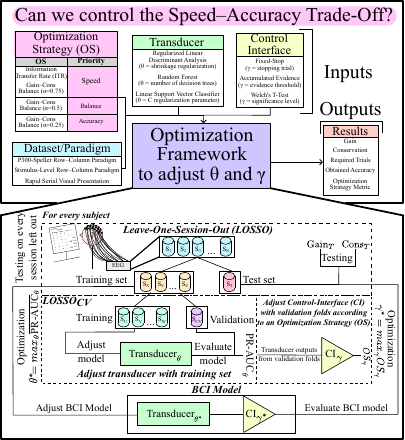}
	\caption{
		The figure illustrates the combined optimization and validation procedure to adjust both transducer and control-interface.
		This process was repeated for every session left out.
		Results on test were averaged across sessions for every subject.
	}
	\label{fig:Val_OptimFW}
\end{figure}

Our framework is a multi-objective optimization procedure (Fig.~\ref{fig:Val_OptimFW}) that jointly optimizes transducers and control-interfaces with the \texttt{optuna} Python package \cite{akiba_optuna_2019}.
The overall process was repeated for every left-out session and results were averaged in terms of Gain and Conservation values.

Within our framework, transducers aim to translate brain signals into evidence and are optimized by maximizing the average \ac{PR-AUC} across validation folds.
\ac{PR-AUC} is a metric particularly suitable for imbalanced problems such as P300-\ac{ERP} detection.
Regarding control-interfaces, their function in this process is dual: (i) to determine the stopping trial at which enough evidence is provided to satisfy an optimization policy and (ii) to assign semantic meaning to transducers' logical outputs as a function of the employed paradigm. 

A \acl{LOSSO} \acl{CV} scheme was adopted.
Within each fold, transducers were tuned using an internal \acl{CV}.
For \ac{SVC}, the regularization parameter $C$ was explored in $[0.001, 1]$.
For \ac{RLDA}, the shrinkage parameter was evaluated in $[0, 1]$, including Ledoit--Wolf regularization \cite{ledoit_well-conditioned_2004}.
For \acl{RF}, the number of estimators was varied between 50 and 300.
Selected models were retrained on the training set and evaluated on the left-out session.

Once evidence is obtained, we define several optimization policies through $\alpha$ in the \acf{GCB}, which weights speed--accuracy contributions:
\begin{equation}
	GCB(\text{Gain}, \text{Cons}, \alpha) = \alpha \text{Gain} + (1-\alpha)\text{Cons},
	\label{eq:gcb}
\end{equation}
where $\alpha \in [0,1]$ determines the prioritization between speed (Gain) and accuracy (Conservation).
Three policies were considered: $\alpha=0.75$ (speed-prioritizing), $\alpha=0.5$ (balanced), and $\alpha=0.25$ (accuracy-prioritizing).
Additionally, the \ac{ITR} was implemented as a reference policy.
The optimization process does not modify the trained classifier; rather, it adjusts the system's stopping criteria to achieve the desired performance in terms of speed and accuracy.

For each defined policy, control-interface parameters were tuned on the training data to satisfy the desired $\alpha$ or bitrate.
Specifically, we employed three different \aclp{ESS} as control-interfaces:
\acl{FS}, which determined the stopping trial;
\acl{AE}, which optimized the evidence threshold;
and \acl{ST}, which adjusted its significance level accordingly.
Performance was summarized in terms of required trials and achieved accuracy.

Experiments combined three classifiers, three \aclp{ESS}, and 63 subjects.
Both \ac{RSVP} and \ac{RCP} paradigms were evaluated.
In \ac{RSVP}, a single stimulus identifies the target, whereas in \ac{RCP} both row and column must be correctly decoded.
To ensure a fair comparison, \ac{RCP} was assessed at both symbol and stimulus levels, we named the last \emph{\acl{RRCP}}.

\section{Results} \label{sec:ress}
\subsection{Qualitative Analysis - Explicit Control of the Speed--Accuracy Trade-Off}
We first provide qualitative evidence illustrating how the proposed framework shapes \ac{BCI} behaviour under different optimization criteria.
All results were obtained following the validation procedure described in Section \ref{meth:val}, evaluating every combination of classifier and \acl{ESS}.
For simplicity, we report here only results from Hoffmann et al.~\cite{hoffmann_efficient_2008}.
Equivalent trends were observed across all experiments and are included within Appendix \ref{appdx}.

\begin{figure}[pos=h]
	\includegraphics[width=\linewidth]{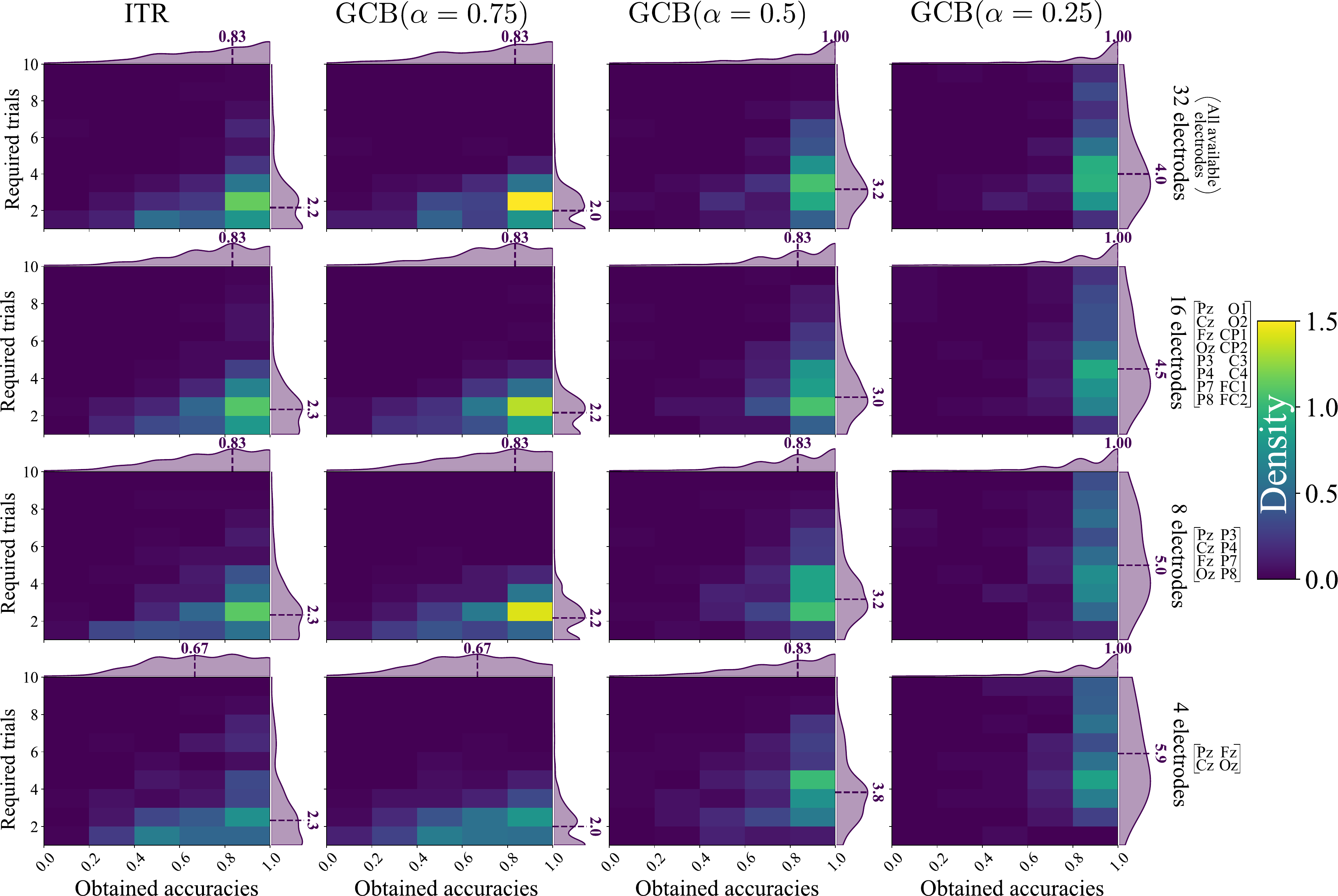}
	\caption{
		Required trials as a function of accuracy for different Hoffmann et al.~electrodes' subsets \cite{hoffmann_efficient_2008}.
		Every histogram was computed from all classifier and \acl{ESS} combinations on each optimization policy.
		Marginal distributions with their respective median are shown on each axis.
		These joint distributions illustrate how different optimization policies shape the operational regions of \acp{BCI} in the speed-accuracy space.
	}
	\label{fig:QualAnal_HoffElSs}
\end{figure}

Fig.~\ref{fig:QualAnal_HoffElSs} presents two-dimensional histograms representing the joint distribution of achieved accuracy and required trials for each optimization policy.
Color intensity reflects the empirical likelihood of observing a given speed--accuracy configuration.
These distributions are obtained by aggregating all subjects, classifiers, and \aclp{ESS}, providing an overall estimate of the system's behavior.
For visualization purposes, the number of trials was truncated to the first 10---where most of the results were observed---and marginal distributions with medians were added.
This truncation does not affect the overall interpretation, since most configurations fall within this range.

Distinct patterns emerge across optimization policies.
\ac{ITR} and GCB($\alpha=0.75$) concentrate probability mass in regions characterized by few trials and broad accuracy dispersion, indicating a preference for faster yet potentially less reliable \acp{BCI}.
Conversely, GCB($\alpha=0.25$) shifts density toward higher accuracies at the expense of requiring more trials.
The GCB($\alpha=0.5$) policy concentrates in the lower-right region of the space, reflecting a balanced compromise between speed and accuracy.
Additionally, reducing the number of electrodes increases density dispersion, which is expected given the loss of discriminative spatial information.

Overall, these results confirm that the proposed \ac{GCB} framework enables explicit modulation of the \acl{SATO}.
While the \ac{ITR} exhibits a clear bias toward speed, \ac{GCB} allows systematic prioritization of either dimension or an intermediate balance.
Furthermore, the density maps provide an empirical estimate of population-level performance distributions, suggesting that new subjects are likely to fall within high-density regions.

\begin{figure*}[pos=h]
	\centering
	\includegraphics[width=\linewidth]{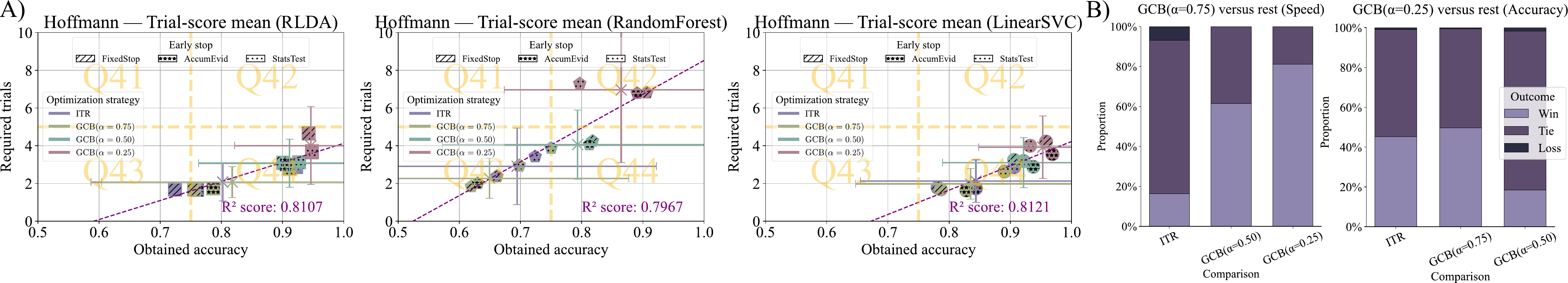}
	\caption{
		(A) Average required trials and obtained accuracies across classifiers, \aclp{ESS}, and optimization policies for Hoffmann et al.~\ac{RSVP} dataset \cite{hoffmann_efficient_2008}; the fitted line illustrates the \acl{SATO}; cross markers indicate the overall mean and standard deviation per optimization policy.
		(B) Proportion of experiments in which each optimization policy won, tied, or lost against others in terms of speed (fewer trials is better, left) or accuracy (higher accuracy is better, right).
		Altogether, these dashboards summarize performance trends at the population level and the relative effectiveness of policies.
	}
	\label{fig:QuanAnal_HoffQuadResults}
\end{figure*}

\subsection{Quantitative Analyses - Robustness of the GCB Framework}
\begin{figure}[pos=h]
	\includegraphics[width=\linewidth]{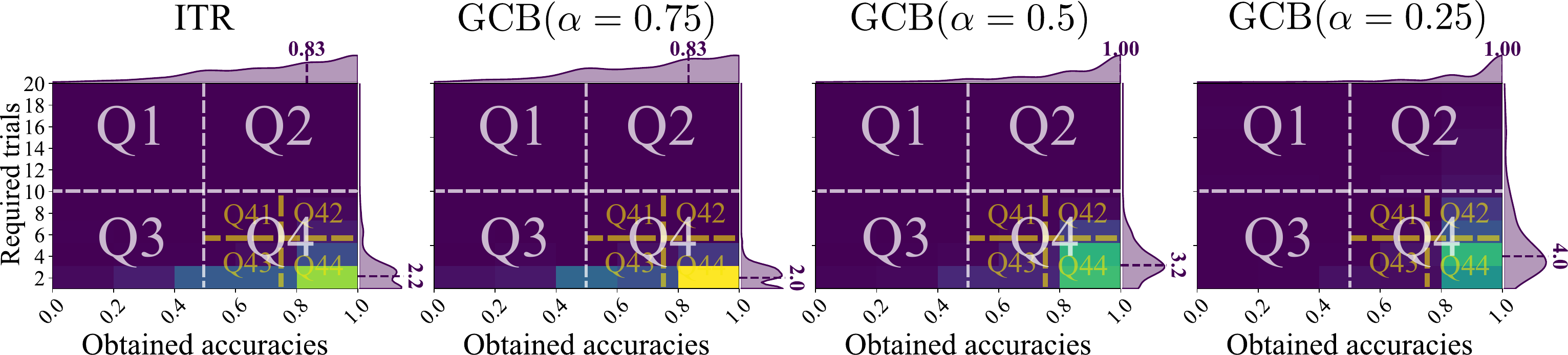}
	\caption{
		Quadrants (Q1-Q4) and subquadrants (Q41-Q44) division example for every optimization policy on Hoffmann et al.~\ac{RSVP} dataset \cite{hoffmann_efficient_2008}.
		This distinction allows for a structured comparison of speed and accuracy profiles across different optimization strategies.
	}
	\label{fig:QuanAnal_QuadExample}
\end{figure}

To compare previous results structurally, we partitioned the speed--accuracy plane into four quadrants as illustrated in Fig.~\ref{fig:QuanAnal_QuadExample}.
Q3 and Q4 correspond to faster \acp{BCI} (fewer trials), while Q2 and Q4 indicate higher accuracy.
Thus, Q4 represents the optimal region, whereas Q1 is the least desirable.

As shown in Fig.~\ref{fig:QuanAnal_QuadExample}, most policies cluster in Q4, suggesting strong overall performance.
To better differentiate them, Q4 was further divided into four analogous subquadrants: Q43 and Q44 denote faster systems, while Q42 and Q44 indicate higher accuracy.

Moreover, the mean and standard deviation of accuracy and required trials were also computed for each classifier, \acl{ESS}, and optimization policy in Fig. \ref{fig:QuanAnal_HoffQuadResults} A).
Consistent with qualitative observations, GCB($\alpha=0.25$) yields higher accuracies but requires more trials spanning across Q42--Q44.
In contrast, GCB($\alpha=0.75$) and \ac{ITR} achieve lower trial counts at the cost of reduced accuracy spanning across Q43--Q44.

Fig. \ref{fig:QuanAnal_HoffQuadResults} A) reveals moderate performance differences across classifiers---particularly for \Acl{RF}---potentially due to lack of regularization.
In contrast, differences across \aclp{ESS} were comparatively small.
Nevertheless, a consistent monotonic relationship between accuracy and required trials is observed, reflecting the intrinsic \acl{SATO}.

These observations are further supported by Fig.~\ref{fig:QuanAnal_HoffQuadResults} B), which quantifies the proportion of experiments in which each optimization policy won, tied, or lost against others under identical conditions (same subject, classifier, \acl{ESS}, and session left-out) by either requiring fewer trials (Fig.~\ref{fig:QuanAnal_HoffQuadResults} B, left) or obtaining higher accuracy (Fig.~\ref{fig:QuanAnal_HoffQuadResults} B, right).
Both \ac{GCB}($\alpha=0.75$) and \ac{ITR} exhibit many ties, consistent with their similar outcomes, while still predominantly winning in terms of speed.
In contrast, \ac{GCB}($\alpha=0.25$) consistently wins in accuracy, confirming its prioritization of high-performance outcomes over trial reduction.

\begin{figure}[pos=h]
	\centering
	\includegraphics[width=\linewidth]{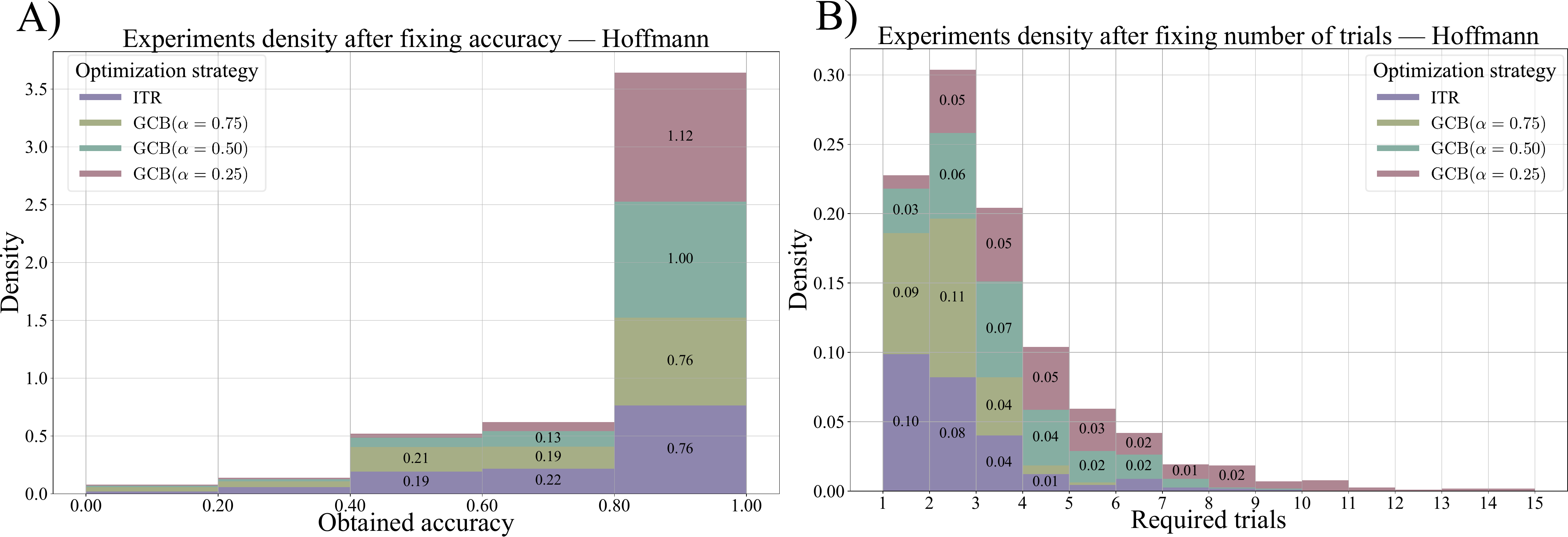}
	\caption{
		Density-stacked histograms from different optimization policies after fixing accuracies or required trials for Hoffmann et al. \cite{hoffmann_efficient_2008}.
		A) Accuracy density for every optimization policy.
		B) Required trials density for every optimization policy.
		These distributions quantify the probability that each policy will achieve specific levels of speed or accuracy.
	}
	\label{fig:CondSBar_Hoff}
\end{figure}

Finally, we also obtained density-stacked histograms (Fig.~\ref{fig:CondSBar_Hoff}) to quantify how likely each policy is to obtain different speed--accuracy qualities.
Particularly, GCB($\alpha=0.25$) exhibits the largest density in the highest accuracy range (Fig.~\ref{fig:CondSBar_Hoff} A).
Conversely, \ac{ITR} and GCB($\alpha=0.75$) concentrate density in the lowest required trials bins (Fig.~\ref{fig:CondSBar_Hoff} B).
These results numerically reinforce that \ac{GCB} effectively conditions \ac{BCI} behaviour according to predefined priorities.

\subsection{Conditional Slices to Select BCIs Under Operational Constraints}
\begin{figure}[pos=h]
	\centering
	\includegraphics[width=\linewidth]{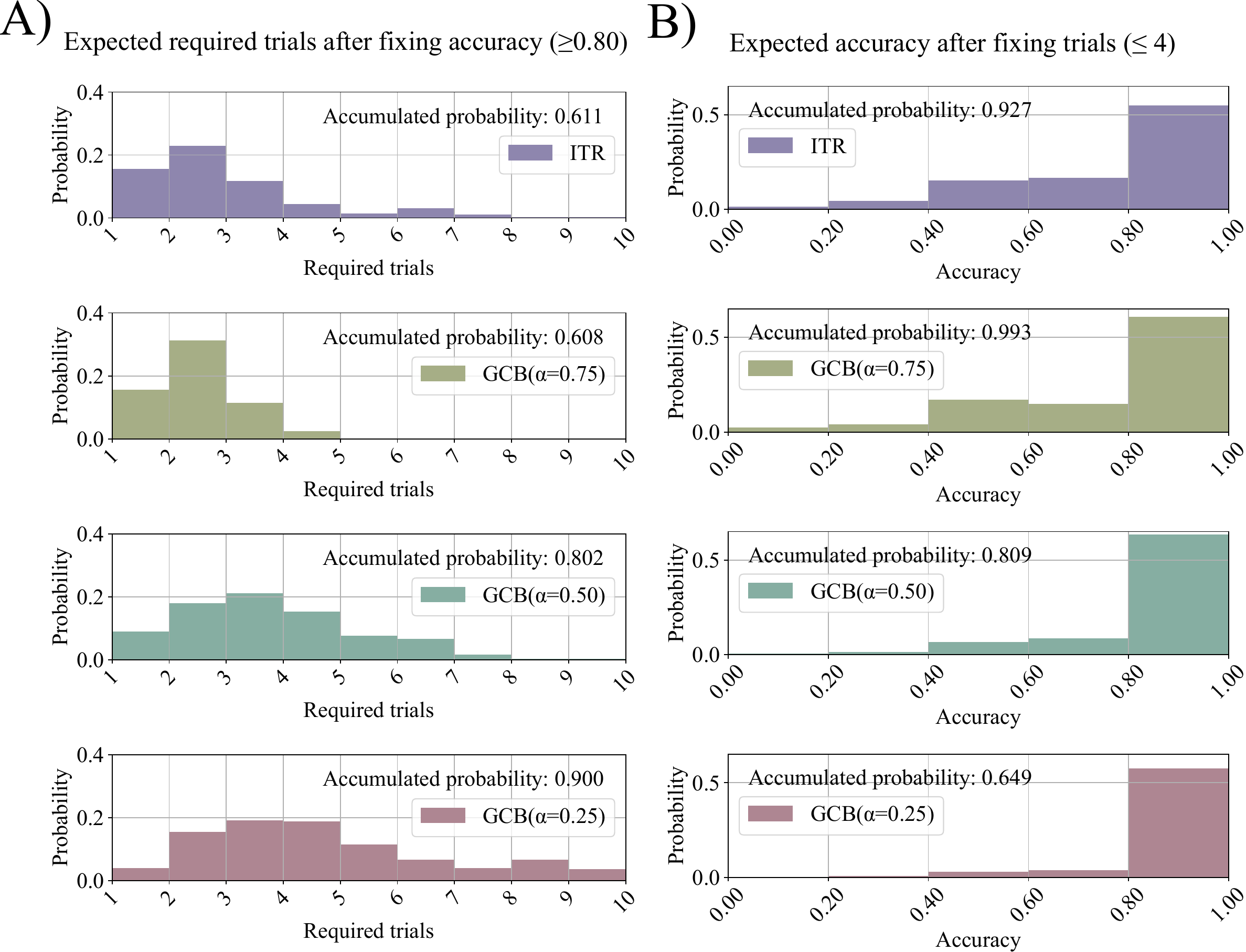}
	\caption{
		Conditional distributions for Hoffmann et al.~\cite{hoffmann_efficient_2008}.
		A) Required trials distribution to obtain, at least, 80\% accuracy.
		B) Accuracy distribution when constraining trials to a maximum of four.
		Please note each policy \emph{accumulated probability} annotation reflects the fraction of the joint distribution considered.
		These conditional distributions allow for the selection of policies under explicit operational constraints.
	}
	\label{fig:CondSlices_Hoff}
\end{figure}

Beyond descriptive statistics, we exploit the joint distributions as maps to derive conditional slices.
These slices estimate expected performance under explicit constraints on accuracy or trial count.
Specifically, they answer three questions: (i) which policy suits different constraints, (ii) how many trials are needed for a target accuracy, and (iii) what accuracy is expected after a fixed number of trials.
Each optimization policy conditional histogram shown in Fig.~\ref{fig:CondSlices_Hoff} was obtained from the joint distribution map by counting samples until reaching a specific level of either accuracy or trials.
Then, sample counts were normalized by the total amount of samples to quantify the observed proportion.

Figure \ref{fig:CondSlices_Hoff} A) presents conditional distributions of required trials to achieve at least 80\% accuracy for each optimization policy.
Crucially, \ac{ITR} and \ac{GCB}($\alpha=0.75$) concentrate most of their probability mass---about 60\%---within the first trials, whereas \ac{GCB}($\alpha=0.25$) accumulates about 90\% of its mass across more trials.
\ac{GCB}($\alpha=0.5$) shows an intermediate trade-off.
Importantly, as just 60\% of the \ac{ITR} and \ac{GCB}($\alpha=0.75$) mass obtains $\ge$80\% accuracy, the remaining 40\% mass falls below 80\% accuracy, which is worth noting in contrast with the remaining 10\% and 20\% mass of \ac{GCB}($\alpha=0.75$) and \ac{GCB}($\alpha=0.5$), respectively.

Similarly, Fig.~\ref{fig:CondSlices_Hoff} B) shows how accuracy distributes after seeing four trials maximum.
Essentially, \ac{ITR} and \ac{GCB}($\alpha=0.75$) distribute 90\% of their overall mass within the 40\%--100\% accuracy range.
In contrast, \ac{GCB}($\alpha=0.5$) and \ac{GCB}($\alpha=0.25$) accumulate almost all of their mass around 80\%--100\% accuracy.
Nonetheless, just 65\% of the \ac{GCB}($\alpha=0.25$) mass is represented, which means that the remaining 35\% is concentrated over higher trial amounts as shown on Fig.~\ref{fig:CondSlices_Hoff} A).

These results demonstrate that the framework enables estimation of performance expectations under operational constraints.
Such maps are particularly useful when minimum accuracy or maximum latency requirements must be satisfied.

\subsection{ITR Reproducibility Analyses Highlight a Bias Toward Speed}
We also employed our Gain--Cons framework to investigate whether the \ac{ITR} inherently prioritizes speed over accuracy.
First, we expressed \ac{ITR} in terms of Gain and Conservation by rewriting required trials and accuracy as functions of these quantities (see Eqs. \ref{eq:gain}, \ref{eq:cons}):
\begin{gather*}
	t^* = t - t \times \text{Gain}, \\
	A(i^*) = A(i) + A(i) \times (\text{Cons} - 1),
\end{gather*}
where $t^*$ denotes the number of required trials and $i^*$ the stopping point of our control-interface.

Then, we included $t^*, A(i^*)$ into Eq. \ref{eq:ITR} by substituing: $P=A(i^*)$ and $\tau=t^* \times \text{trial duration}$.
Additionally, we simulated all possible stopping conditions to study how the \ac{ITR} changes as a function of $t^*, i^*$, which depend on $t, A(i)$, Gain and Conservation.
As these baseline values, $t$ and $A(i)$, are fixed---corresponding to the maximum number of trials and accuracy, respectively---the \ac{ITR} only changed according to Gain and Conservation.
This means that any variation in the \ac{ITR} is due solely to a change in the system's operating point, and not to modifications to the classifier.

\begin{figure}[pos=h]
	\centering
	\includegraphics[width=\linewidth]{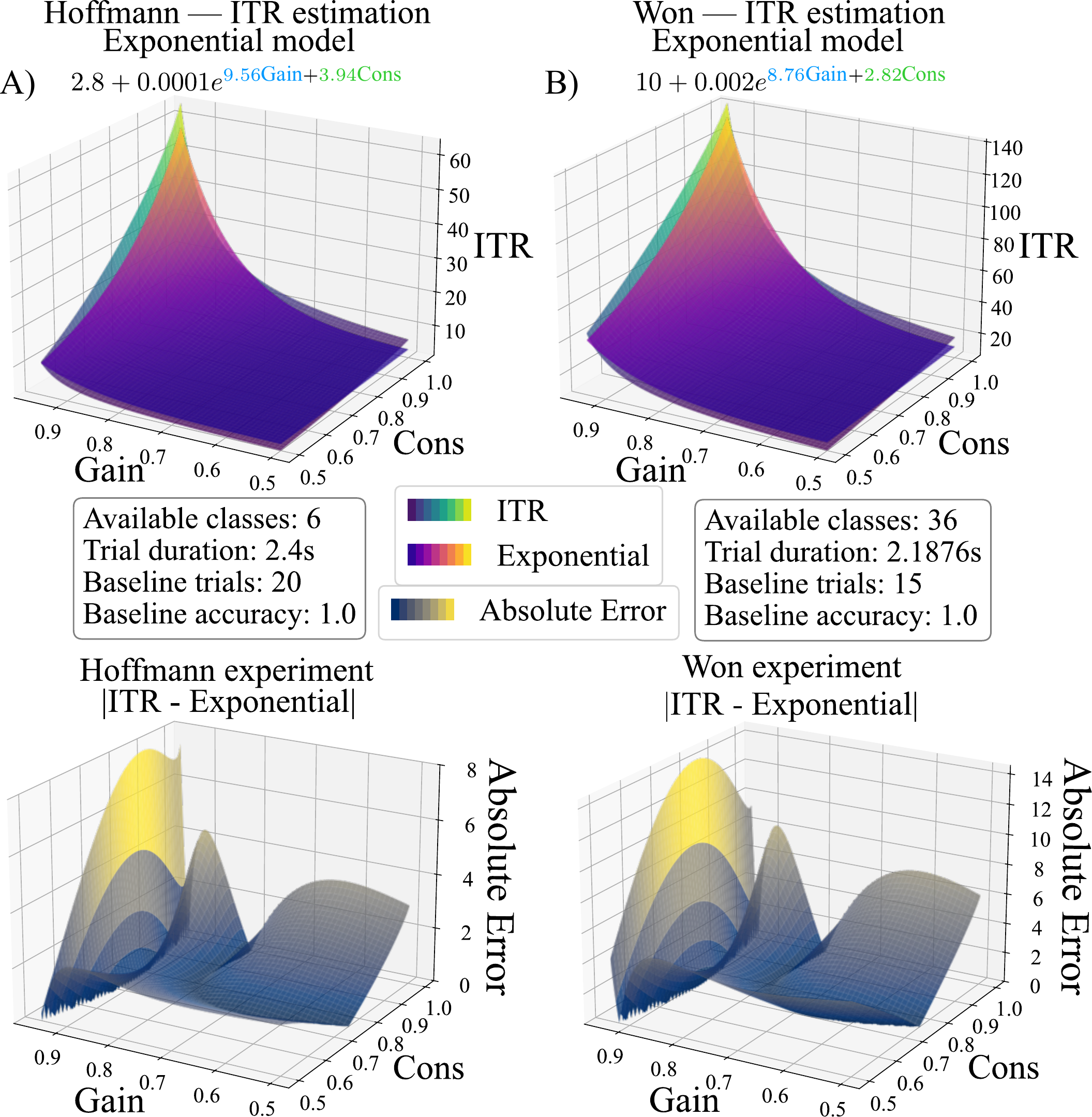}
	\caption{
		\ac{ITR} estimated model and absolute error across different experimental parameters.
		A) Results for Hoffmann et al.~\cite{hoffmann_efficient_2008} parameters.
		B) Results for Won et al.~\cite{won_eeg_2022} parameters.
		The fitted models illustrate the sensitivity of the \ac{ITR} to variations in speed and accuracy.
	}
	\label{fig:GCB-ITR_estim}
\end{figure}

\ac{ITR} values were then compared against a non-linear model following the form: $d + c\exp(Gain^a + Cons^b)$ as shown in Fig.~\ref{fig:GCB-ITR_estim}.
Our analysis focused within the region of interest: Gain, Cons $\in [0.5, 1]$ which corresponds to fast and accurate decisions.
Parameters were estimated through non-linear least squares.

The estimated non-linear models explained 96.17\% and 97.33\% of the \ac{ITR} variance ($R^2$) for Hoffmann et al.~\cite{hoffmann_efficient_2008} and Won et al.~\cite{won_eeg_2022} datasets, respectively.
Importantly, fitted coefficients consistently assigned greater weight to Gain in comparison to Cons.
This indicates that \ac{ITR} increases more steeply with speed gains than with equivalent accuracy conservation improvements.

\begin{figure}[pos=h]
	\includegraphics[width=\linewidth]{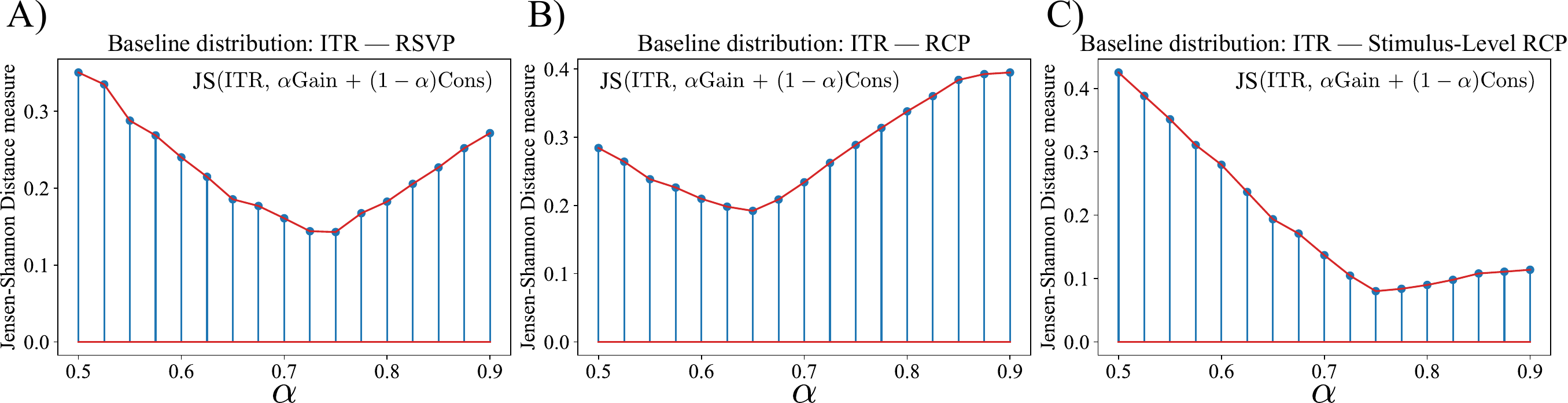}
	\caption{
		Jensen--Shannon distance between \acp{BCI} optimized with the \ac{ITR} and \acp{BCI} adjusted with different \ac{GCB}'s $\alpha$ across \ac{RSVP} (A), \ac{RCP} (B), and \acl{RRCP} (C) paradigms.
		The closer to zero, the more similar and vice versa.
		This analysis quantifies the empirical similarity between systems optimized using the \ac{ITR} and those using the \ac{GCB} at different $\alpha$ values, identifying the range of $\alpha$ values that best aligns with \ac{ITR} optimization.
	}
	\label{fig:GCB-ITR_AlphaSim}
\end{figure}

Additionally, in a complementary analysis, we empirically adjusted all classifiers and \aclp{ESS} using the optimization framework described in Section \ref{meth:val}.
Optimization policies varied Eq. \ref{eq:gcb} $\alpha$ within the interval $[0.5, 1]$ so as to compare similarities between \ac{ITR} and different $\alpha$ values in \ac{GCB} when comparing \acp{BCI}.
Similarity was assessed using the Jensen--Shannon distance (Fig.~\ref{fig:GCB-ITR_AlphaSim}) as it allows us to rank similarities between distributions contrary to other measures such as the Kullback-Leibler divergence.

Minimum distances---maximum similarities---were consistently observed for $\alpha > 0.5$.
For \ac{RSVP} and \acl{RRCP}, peak similarity occurred at $\alpha \in [0.7, 0.8]$ (Figs.~\ref{fig:GCB-ITR_AlphaSim} A, C), whereas \ac{RCP} peak similarity shifted slightly toward lower $\alpha$ (Fig.~\ref{fig:GCB-ITR_AlphaSim} B).
This shift toward accuracy-conservative policies was attributed to \ac{RCP}'s conjunctive constraint, where both row and column need to be predicted correctly to decode a symbol---leading to lower accuracies when employing faster \acp{BCI}.

Overall, both theoretical reconstruction and empirical similarity analyses support the conclusion that \ac{ITR} inherently favours faster \acp{BCI}.

\section{Discussion} \label{sec:disc}
This section interprets and contextualizes the results obtained, analyzing their methodological and practical implications for the design and evaluation of \ac{BCI} systems.
In particular, we discuss how the proposed framework allows for explicit control of the \acl{SATO} (Section \ref{disc:sato-control}), how this behaviour holds consistently across different classifiers and \aclp{ESS} (Section \ref{disc:consistency}), and how joint and conditional distributions facilitate the selection of configurations under operational constraints (Section \ref{disc:cond-perf}).
Additionally, we examine the implications of the speed bias inherent in the \ac{ITR} metric, contrasting it with the explicit control capability introduced by the Gain--Conservation approach (Section \ref{disc:ITR-bias}).
Finally, we also discuss the main limitations of this work and outline future research directions (Section \ref{disc:limits}).

\subsection{Explicit Control of the Speed--Accuracy Trade-Off} \label{disc:sato-control}
The results demonstrate that the proposed \ac{GCB} framework enables explicit and continuous control over the \acl{SATO} through the parameter $\alpha$.
By construction, $\alpha$ directly modulates the relative importance of speed and accuracy, allowing the optimization process to target different operational regimes.
In particular, larger values of $\alpha$ favour faster decisions by prioritizing temporal efficiency, whereas smaller values emphasize classification reliability.
This behaviour is consistently observed across datasets and paradigms, confirming that the framework provides a predictable and interpretable mechanism to navigate the trade-off.

In particular, speed-prioritizing configurations can be achieved by optimizing \aclp{ESS} using either the \ac{ITR}---which inherently biases performance toward shorter decision times---or \ac{GCB} with high $\alpha$ values (e.g., $\alpha=0.75$).
Conversely, accuracy-prioritizing operation emerges when using low $\alpha$ values (e.g., $\alpha=0.25$), where the optimization favours more conservative decisions at the expense of speed.
Intermediate settings (e.g., $\alpha=0.5$) provide a balanced regime, yielding solutions that neither excessively prioritize speed nor accuracy.
Importantly, this continuum of behaviours is obtained within a single unified formulation, avoiding the need to switch between different measures.

These findings have direct implications for the design of application-specific \ac{BCI} systems.
Tasks such as speller-based communication, where throughput is critical but moderate error rates can be tolerated and corrected, may benefit from speed-prioritizing or balanced configurations (i.e., $\alpha \ge 0.5$).
In contrast, safety-critical applications such as wheelchair or robotic control require highly reliable decisions, making accuracy-prioritizing settings (i.e., $\alpha$ \textless 0.5) more appropriate.
This flexibility is particularly relevant given the diversity of modern \ac{BCI} applications, ranging from assistive communication to online control systems.

Moreover, the density maps (Fig.~\ref{fig:QualAnal_HoffElSs}) provide additional insight into the distribution of achievable performance across subjects.
High-density regions indicate combinations of speed and accuracy that are consistently attainable, suggesting that new users are likely to operate within these regions.
This observation supports the practical relevance of the proposed framework, as it not only enables explicit trade-off control but also offers a data-driven perspective on expected population-level performance.
These regions are derived from the aggregation of all subjects and configurations, providing an overall estimate of expected behavior.

On balance, these results confirm that, unlike traditional metrics such as \ac{ITR}, which implicitly favour speed, the proposed \ac{GCB} formulation exposes the \acl{SATO} as an explicit and tunable design variable termed $\alpha$.
This facilitates principled optimization and more transparent alignment between system behaviour and application requirements.

\subsection{Consistency Across Classifiers and Early-Stopping Strategies} \label{disc:consistency}
Average performance across classifiers and \aclp{ESS} was largely consistent, indicating that most configurations behave similarly under the same optimization policy.

Differences across classifiers and \aclp{ESS} were comparatively small, suggesting that optimization criteria play a more dominant role than specific model components when meeting speed and accuracy requirements.
However, \Acl{RF} exhibited systematically worse performance, yielding lower accuracies and requiring more trials than \ac{RLDA} and \ac{SVC}.
This behaviour is likely due to the lack of regularization, which is critical in \ac{ERP}-based \ac{BCI} systems where high-dimensional features and limited samples can easily lead to overfitting.

Despite not explicitly regularizing \Acl{RF}, the proposed Gain--Cons framework was able to reveal these performance differences when compared to linear and regularized methods.
This highlights the utility of the framework as a diagnostic tool to detect suboptimal or unstable configurations.

Additionally, models in Fig.~\ref{fig:QuanAnal_HoffQuadResults} A) follow a monotonic trend that can be approximated by a linear relationship between accuracy and required trials.
This trend reflects that the intrinsic \acl{SATO} can be controlled as a function of $\alpha$, turning this trade-off into a controllable design variable.

Generally, although different classifiers and \aclp{ESS} affect performance, optimization policies such as \ac{ITR} and \ac{GCB} play a central role in determining when to stop data acquisition.
In practice, these policies define whether a system should operate in a faster or more reliable regime depending on application constraints.
Therefore, incorporating an explicit optimization component is essential to meet operational requirements in real-world \acp{BCI}.

\subsection{Operational Selection Through Conditional Performance Maps} \label{disc:cond-perf}
The proposed Gain--Cons framework enables the identification of suitable \ac{BCI} operating points under explicit operational constraints.
By analyzing conditional slices (Fig.~\ref{fig:CondSlices_Hoff}) derived from joint speed--accuracy distributions (Fig.~\ref{fig:QualAnal_HoffElSs}), it becomes possible to directly compare optimization policies while enforcing application-specific requirements.

In particular, conditional performance maps provide a principled mechanism to evaluate policies under constraints such as a maximum number of trials, a minimum required accuracy, or a combination of both.
Conditioning on the number of trials yields a probability distribution over achievable accuracies, while conditioning on accuracy thresholds provides a distribution over the required number of trials.
This perspective enables a direct assessment of how reliably a given policy satisfies timing or performance constraints.

The distributions shown in Fig.~\ref{fig:CondSlices_Hoff} illustrate clear differences across optimization strategies.
Speed-prioritizing policies, such as \ac{ITR} or \ac{GCB}($\alpha=0.75$), concentrate probability mass in early trials but exhibit wider variability in accuracy.
In contrast, accuracy-prioritizing configurations, such as \ac{GCB}($\alpha=0.25$), concentrate probability mass near high accuracies, at the cost of increased number of trials.
Balanced configurations distribute probability more evenly, offering intermediate trade-offs between reliability and speed.

These properties make conditional performance maps particularly valuable for application-driven design.
For instance, safety-critical systems such as wheelchair control require highly reliable decisions, favouring policies whose probability mass concentrates on near-perfect accuracy, even if additional trials are needed.
Conversely, communication-prioritizing applications such as \ac{BCI} spellers may prioritize faster interaction under moderate accuracy constraints, making speed-prioritizing or balanced policies more suitable.
Rather than relying on aggregate metrics alone, these maps allow selecting policies based on their likelihood of satisfying specific operational requirements.

Beyond policy selection, conditional maps also provide a diagnostic tool to analyze system behaviour.
They can reveal whether an optimization policy meets expected performance targets and highlight variability across subjects.
Furthermore, they can guide system design by identifying regions of the speed--accuracy space that are both achievable and desirable, informing the joint tuning of transducers and control-interfaces.

As a whole, conditional performance maps extend traditional evaluation by introducing a probabilistic and constraint-aware perspective.
They enable informed decision-making under realistic operating conditions and complement the explicit \acl{SATO} control provided by the \ac{GCB} framework.

\subsection{ITR Exhibits an Intrinsic Bias Toward Speed} \label{disc:ITR-bias}
Both theoretical and empirical analyses indicate that \ac{ITR} inherently prioritizes speed over accuracy.
Its non-linear reconstruction (Fig.~\ref{fig:GCB-ITR_estim}) shows strong sensitivity to speed gains, consistently weighting Gain more than Conservation.
This asymmetry is supported by Jensen--Shannon similarity results (Fig.~\ref{fig:GCB-ITR_AlphaSim}), where \ac{ITR}-optimized systems align most closely with speed-prioritizing \ac{GCB} configurations ($\alpha \in [0.6, 0.8]$), confirming its preference for fast operating regimes.

These findings have important implications for \ac{BCI} evaluation.
Because \ac{ITR} implicitly favours rapid decisions, relying solely on this metric may disadvantage systems designed for high reliability.
Evaluation should therefore reflect the intended application: while \ac{ITR} suits communication scenarios requiring high throughput, it is less appropriate for safety-critical contexts that demand near-perfect accuracy.

This bias does not invalidate \ac{ITR} as a performance measure; it remains effective when speed is the primary objective, such as in \ac{BCI} spellers.
However, its prioritization should be explicitly acknowledged when interpreting results or designing optimization strategies.
In contrast, applications like wheelchair or robotic control require more conservative criteria that emphasize accuracy over speed.

The \ac{ITR} bias also depends on paradigm-specific constraints.
For example, in \ac{RCP} paradigms, the need to correctly identify row and column favours conservative decisions, shifting similarities between \ac{GCB}($\alpha$) and \ac{ITR} toward lower $\alpha$ values.
Nevertheless, \ac{ITR} consistently promotes faster decisions.
From a broader perspective, it can be seen as an implicit speed-weighted trade-off within the Gain--Cons space, however, \ac{GCB} provides a linear and explicit control of this balance, enabling more predictable behaviour across paradigms.

\subsection{Limitations} \label{disc:limits}
This study has focused on \ac{ERP} paradigms based on discrete stimuli and synchronous decision-making scenarios, where the number of trials constitutes an explicit control variable.
Although the proposed framework is conceptually independent of the classifier and the paradigm, its experimental validation has focused on P300-\ac{ERP} paradigms and it does not cover asynchronous \acp{BCI}, non-\ac{ERP} signals, or other \acp{ERP} such as \acp{SSVEP}.
These modalities require more careful consideration as the definition of speed and accuracy may need alternative formulations.

The use of a \acl{bBCI} influences the relative quantification of Gain and Conservation, although this approach is suitable for comparing configurations within the same system and can be refined as models improve.
This limitation becomes more pronounced when comparing highly heterogeneous systems, where the baseline definition more strongly shapes the quantitative interpretation of results.

Finally, results were obtained in an offline validation setting; therefore, their performance in real-time \ac{BCI} systems could be affected by additional factors such as processing latencies or user adaptation.
These considerations delimit the current scope of the framework without compromising its methodological generality.

\section{Conclusion} \label{sec:conc}
In this work, we introduced the \acl{GCB} framework as a principled approach to jointly analyze and explicitly control \acp{BCI} in terms of speed and accuracy.
By optimizing transducers and control-interfaces under different policies and evaluating the framework on established P300-based paradigms, such as \ac{RSVP} \cite{hoffmann_efficient_2008} and \ac{RCP} \cite{won_eeg_2022}, we enabled systematic exploration of the joint speed--accuracy space and the derivation of conditional distributions describing expected performance under fixed accuracy or latency constraints.

Our results show that the widely used \ac{ITR} exhibits an inherent bias toward faster systems, often at the expense of accuracy.
In contrast, the proposed \acl{GCB} formulation introduces an interpretable trade-off parameter $\alpha$ that makes this compromise explicit, allowing controlled prioritization of speed, accuracy, or intermediate regimes.
This transforms the \acl{SATO} from an implicit limitation into a tunable design variable.
Importantly, adaptation could be achieved without online retraining or further modification of the classifier, as only the control policy ($\alpha$) needs to be adjusted, which is significantly more practical for real-world deployment.

Furthermore, by relying on speed--accuracy density maps and conditional distributions rather than simple averages, the framework enables anticipating the impact of policy changes.
This predictive capability is essential for context-aware \acp{BCI}, understood as systems that dynamically adjust their operating point depending on task demands, user characteristics, or recent interaction history.

Overall, the proposed framework provides a general and robust methodology for analyzing, comparing, and tuning \acp{BCI} under application-specific requirements, enabling subject-level estimation of achievable performance for new users, principled comparison of different \aclp{BCIm}, and informed selection of transducers and control-interfaces.
In this sense, it may offer a practical foundation for adaptive \ac{BCI} systems that can adjust their behaviour without retraining and anticipate the consequences of such adaptations.



\printcredits

\subsection*{Funding sources}
This work was supported by the Predoctoral Research Grants of the Universidad Autónoma de Madrid (FPI-UAM) and by PID2023-149669NB-I00 (MCIN/AEI and ERDF – "A way of making Europe").

\subsection*{Declaration of competing interest}
The authors declare that they have no known competing financial interests or personal relationships that could have appeared to influence the work reported in this paper.

\subsection*{Code availability}
The published code to reproduce this article can be found on \href{https://github.com/GNB-UAM/P300-BCI-OptFramework.git}{our research group GitHub page}.

\appendix
\section{Appendix} \label{appdx}
This appendix complements Section~\ref{sec:ress} by reporting additional results on the Won et al.~\ac{RCP} dataset \cite{won_eeg_2022} and the derived \acl{RRCP} modality.
The separation from the main manuscript was intentionally introduced to avoid redundancy, as all evaluated modalities consistently reflected the same central observation: the proposed framework enables control of the \acl{SATO} through the $\alpha$ parameter of \ac{GCB}($\alpha$), while subject performance maps and operational constraints provide insight into the impact of policy changes for context-aware \acp{BCI}.

Overall, the obtained conclusions remain consistent with those discussed for the Hoffmann et al.~\ac{RSVP} dataset \cite{hoffmann_efficient_2008}.
The main exception lies in the original \ac{RCP} modality, whose conjunctive decoding constraint requires simultaneously identifying row and column P300 responses, thereby favouring more accuracy-conservative solutions and reducing the similarity between \ac{GCB}($\alpha=0.75$) and \ac{ITR}.
For clarity, the appendix is divided into two subsections discussing the \ac{RCP} (Section \ref{appdx:rcp-ress}) and \acl{RRCP} (Section \ref{appdx:rrcp-ress}) modalities, respectively.

\subsection{Framework effects on the Row--Column Paradigm} \label{appdx:rcp-ress}
First, it is important to remark the inherent conjunctive decoding constraint of the \ac{RCP} modality, where both row and column predictions must be simultaneously correct to count as a successful decoding.
This constraint necessarily encourages the \ac{RCP} modality to favour more accuracy-conservative solutions than the Hoffmann et al.~\ac{RSVP} paradigm \cite{hoffmann_efficient_2008} discussed throughout the main manuscript.

\begin{figure*}[pos=h]
	\centering
	\includegraphics[width=\linewidth]{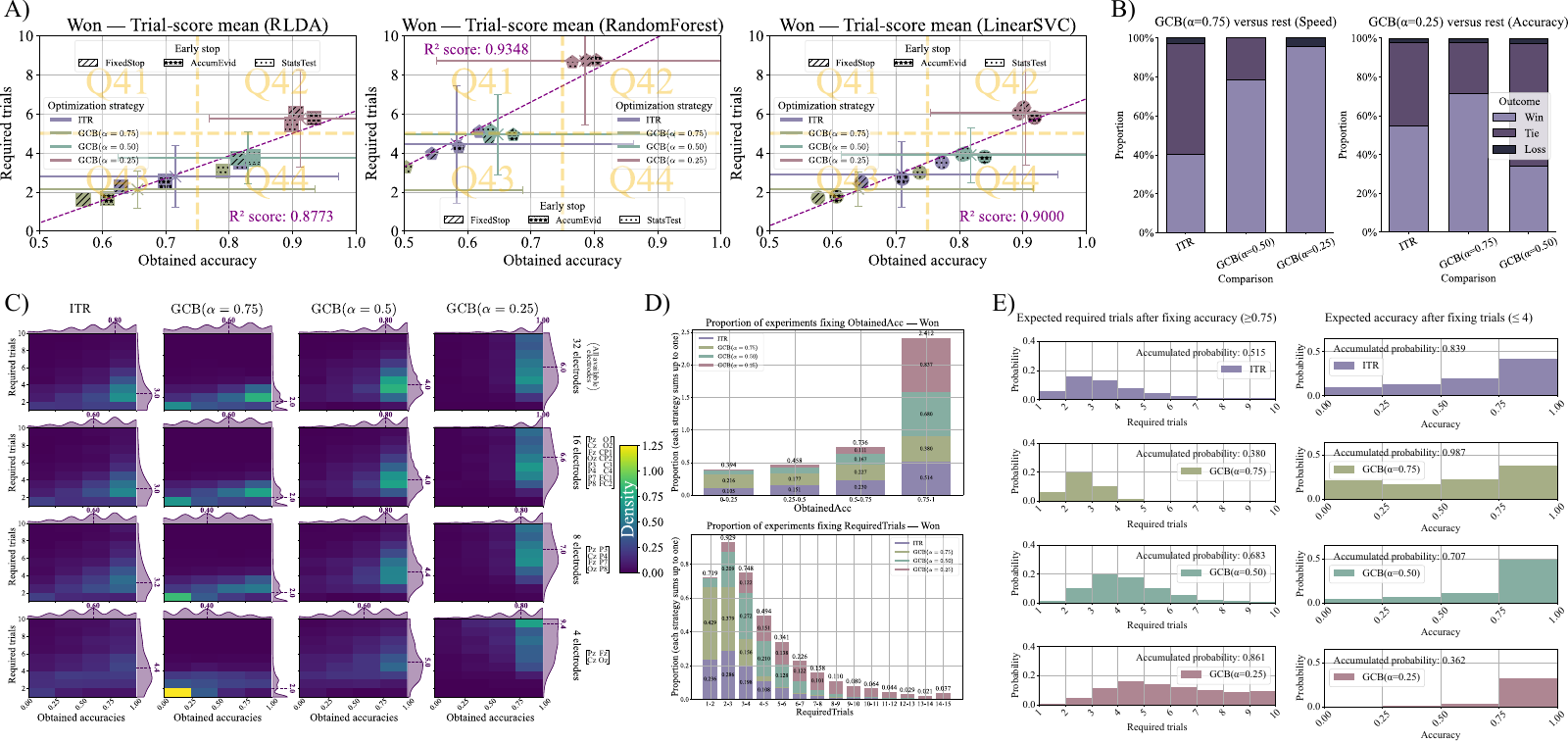}
	\caption{
		Reproduced results on the \ac{RCP} modality from Won et al.~dataset \cite{won_eeg_2022}.
		(A) Average required trials and obtained accuracies across classifiers, \aclp{ESS}, and optimization policies; the fitted line illustrates the \acl{SATO}; cross markers indicate the overall mean and standard deviation per optimization policy.
		(B) Proportion of experiments in which each optimization policy won, tied, or lost against others in terms of speed (fewer trials is better, left) or accuracy (higher accuracy is better, right).
		(C) Required trials as a function of accuracy for different electrodes' subsets.
		Every histogram was computed from all classifier and \acl{ESS} combinations on each optimization policy.
		(D) Density-stacked histograms from different optimization policies after fixing either accuracy (top) or required trials (bottom).
		(E) Conditional distributions after constraining either a minimum of 75\% accuracy (left) or a maximum of four trials (right).
	}
	\label{fig:Appdx_RCP_Results}
\end{figure*}

We applied the Won et al.~\ac{RCP} modality \cite{won_eeg_2022} to our optimization framework and compiled all results in Fig.~\ref{fig:Appdx_RCP_Results}.
Overall, this figure reproduces the same general trends and characteristics previously observed for the Hoffmann et al.~\ac{RSVP} modality \cite{hoffmann_efficient_2008}.

In particular, Fig.~\ref{fig:Appdx_RCP_Results} A) highlights the ability of the framework to navigate the speed--accuracy trade-off within the \ac{RCP} modality through the proposed Gain--Cons Balance.
More specifically, the $\alpha$ parameter enables selecting lower values for accuracy-sensitive strategies and higher values for faster \acp{BCI}.
Figure~\ref{fig:Appdx_RCP_Results} B) compares the proportion of times each optimization policy won, lost, or tied against the others in terms of speed and accuracy.
Additionally, Figs.~\ref{fig:Appdx_RCP_Results} C), D), and E) illustrate the impact of policy changes through subject performance maps and operational constraints.

Although these results closely resemble those obtained for the Hoffmann et al.~\ac{RSVP} modality \cite{hoffmann_efficient_2008}, a relevant difference emerges in the similarity between \ac{GCB}($\alpha=0.75$) and \ac{ITR} policies shown in Fig.~\ref{fig:Appdx_RCP_Results} B).
Specifically, Fig.~\ref{fig:Appdx_RCP_Results} B) shows that \ac{GCB}($\alpha=0.75$) ties with \ac{ITR} in nearly 50\% of the experiments within the \ac{RCP} modality, whereas these policies tied in more than 70\% of the experiments for the \ac{RSVP} modality shown in Fig.~\ref{fig:QuanAnal_HoffQuadResults} B).
This behaviour was further confirmed in Fig.~\ref{fig:GCB-ITR_AlphaSim}, where differences between \ac{GCB}($\alpha=0.75$) and \ac{ITR} policies increased notably only for the \ac{RCP} modality shown in Fig.~\ref{fig:GCB-ITR_AlphaSim} B).

Overall, the \ac{RCP} results remain equivalent to those obtained for the \ac{RSVP} modality except for the reduced similarity between \ac{GCB}($\alpha=0.75$) and \ac{ITR}.
We attribute this difference to the non-linear evolution of \ac{ITR} shown in Fig.~\ref{fig:GCB-ITR_estim} B), which makes the \ac{ITR} look out for higher accuracies at the expense of requiring more trials.
Ultimately, this behaviour is a direct consequence of the conjunctive decoding constraint inherent to the \ac{RCP} modality, which implies that the structure of the paradigm can directly influence the \acl{SATO} induced by the metric.

\subsection{Framework effects on the Stimulus-Level modality of Row--Column Paradigm} \label{appdx:rrcp-ress}
Contrary to the conjunctive decoding constraint inherent to the original \ac{RCP} modality, the \acl{RRCP} modality derived from the Won et al. experiment \cite{won_eeg_2022} only requires determining whether a P300-\ac{ERP} response occurred.
This formulation removes the need to simultaneously predict both rows and columns correctly, thereby increasing the likelihood of achieving higher accuracies with fewer trials.

\begin{figure*}[pos=h]
	\centering
	\includegraphics[width=\linewidth]{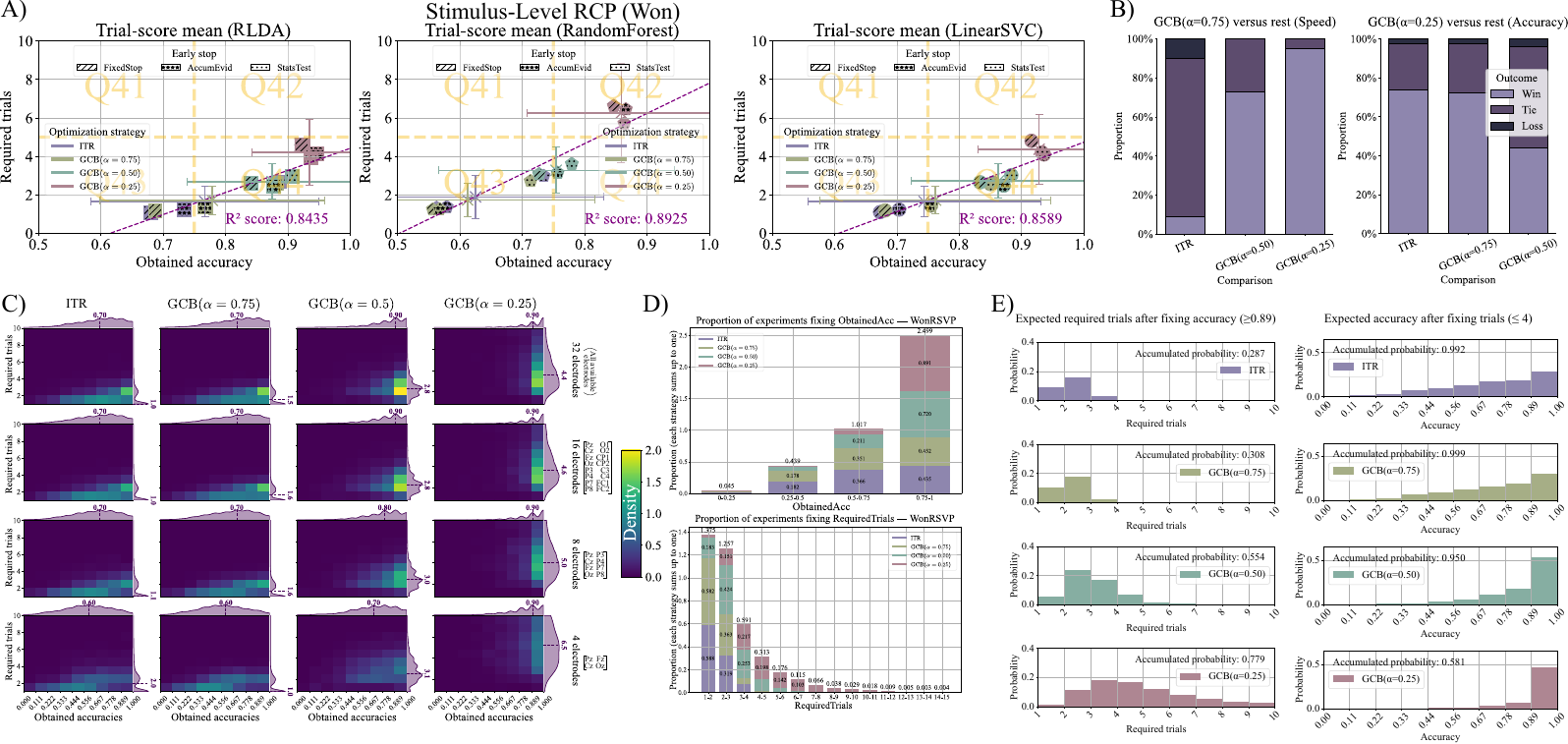}
	\caption{
		Reproduced results on the \acl{RRCP} modality from Won et al.~dataset \cite{won_eeg_2022}.
		(A) Average required trials and obtained accuracies across classifiers, \aclp{ESS}, and optimization policies; the fitted line illustrates the \acl{SATO}; cross markers indicate the overall mean and standard deviation per optimization policy.
		(B) Proportion of experiments in which each optimization policy won, tied, or lost against others in terms of speed (fewer trials is better, left) or accuracy (higher accuracy is better, right).
		(C) Required trials as a function of accuracy for different electrodes' subsets.
		Every histogram was computed from all classifier and \acl{ESS} combinations on each optimization policy.
		(D) Density-stacked histograms from different optimization policies after fixing either accuracy (top) or required trials (bottom).
		(E) Conditional distributions after constraining either a minimum of 89\% accuracy (left) or a maximum of four trials (right).
	}
	\label{fig:Appdx_RRCP_Results}
\end{figure*}

We applied the \acl{RRCP} modality to our optimization framework and compiled all results in Fig.~\ref{fig:Appdx_RRCP_Results}.
Overall, this figure reproduces the same trends and characteristics previously observed for the other evaluated modalities.

In particular, the obtained results remain highly consistent with those previously reported for the \ac{RSVP} modality.
Unlike the original \ac{RCP} modality, we did not observe substantial differences between \ac{GCB}($\alpha=0.75$) and \ac{ITR}.
We attribute this behaviour to the removal of the conjunctive decoding constraint inherent to the original \ac{RCP} formulation.
By eliminating this constraint, the \acl{RRCP} modality directly increases achievable accuracies while simultaneously reducing the number of required trials.
Consequently, the resulting optimization behaviour becomes considerably more similar to that observed for the \ac{RSVP} modality discussed throughout the main manuscript.
This supports the hypothesis that the differences observed in \acp{RCP} are primarily due to structural limitations of the paradigm rather than to the behavior of the proposed framework.


\bibliographystyle{cas-model2-names}
\bibliography{Bibliography/References}

\end{document}